\documentclass[10pt, conference]{IEEEtran}

\usepackage{amsmath,amsfonts}
\usepackage{algorithmic}
\usepackage{graphicx}
\usepackage{textcomp}
\usepackage{xcolor}
\usepackage{diagbox}
\usepackage{multirow}
\usepackage{booktabs}
\usepackage{listings}
\usepackage{makecell}
\usepackage{soul}
\usepackage{colortbl}
\usepackage{scrextend}
\usepackage{array}
\usepackage{hyperref}
\usepackage{subfigure}
\usepackage{fontawesome5}
\usepackage{cite}

\newcommand{\estimates}{\overset{\scriptscriptstyle\wedge}{=}}

\newcommand{\question}[1]{%
    \medskip%
    \noindent\fcolorbox{black}{blue!05}{%
        \parbox{0.97\linewidth}{%
            \textbf{\faQuestionCircle\;RQ} #1%
        }%
    }%
    \medskip%
}%

\newcommand{\answer}[1]{%
    \medskip%
    \noindent\fcolorbox{black}{green!05}{%
        \parbox{0.97\linewidth}{%
            \textbf{\faExclamationCircle\;RQ} #1%
        }%
    }%
    \medskip%
}%

\definecolor{delim}{RGB}{20,105,176}
\definecolor{numb}{RGB}{106, 109, 32}
\definecolor{string}{rgb}{0.64,0.08,0.08}

\lstdefinelanguage{json}{
    numbers=left,
    numberstyle=\small,
    rulecolor=\color{black},
    showspaces=false,
    showtabs=false,
    breaklines=true,
    postbreak=\raisebox{0ex}[0ex][0ex]{\ensuremath{\color{gray}\hookrightarrow\space}},
    breakatwhitespace=true,
    basicstyle=\ttfamily\small,
    upquote=true,
    morestring=[b]",
    stringstyle=\color{string},
    literate=
     *{0}{{{\color{numb}0}}}{1}
      {1}{{{\color{numb}1}}}{1}
      {2}{{{\color{numb}2}}}{1}
      {3}{{{\color{numb}3}}}{1}
      {4}{{{\color{numb}4}}}{1}
      {5}{{{\color{numb}5}}}{1}
      {6}{{{\color{numb}6}}}{1}
      {7}{{{\color{numb}7}}}{1}
      {8}{{{\color{numb}8}}}{1}
      {9}{{{\color{numb}9}}}{1}
      {\{}{{{\color{delim}{\{}}}}{1}
      {\}}{{{\color{delim}{\}}}}}{1}
      {[}{{{\color{delim}{[}}}}{1}
      {]}{{{\color{delim}{]}}}}{1},
}

\begin{document}

\title{How Dataflow Diagrams Impact Software Security Analysis: an Empirical Experiment}

\author{
    \IEEEauthorblockN{Simon Schneider\IEEEauthorrefmark{1}, Nicolás E. Díaz Ferreyra\IEEEauthorrefmark{1}, Pierre-Jean Quéval\IEEEauthorrefmark{2}, Georg Simhandl\IEEEauthorrefmark{2}, \\Uwe Zdun\IEEEauthorrefmark{2}, Riccardo Scandariato\IEEEauthorrefmark{1}}
    
    \IEEEauthorblockA{\IEEEauthorrefmark{1}Hamburg University of Technology, \emph{firstname.lastname@tuhh.de}}
    
    \IEEEauthorblockA{\IEEEauthorrefmark{2}University of Vienna, \emph{firstname.lastname@univie.ac.at}}
}

\markboth{SANER 2024}{Schneider \MakeLowercase{\textit{et al.}}: How Architectural Models and Traceability Links to Code Impact Software Security Analysis Activities: an Empirical Experiment}

\maketitle

\begin{abstract}
Models of software systems are used throughout the software development lifecycle. 
Dataflow diagrams (DFDs), in particular, are well-established resources for security analysis. 
Many techniques, such as threat modelling, are based on DFDs of the analysed application. 
However, their impact on the performance of analysts in a security analysis setting has not been explored before.
In this paper, we present the findings of an empirical experiment conducted to investigate this effect.
Following a within-groups design, participants were asked to solve security-relevant tasks for a given microservice application.
In the control condition, the participants had to examine the source code manually.
In the model-supported condition, they were additionally provided a DFD of the analysed application and traceability information linking model items to artefacts in source code.
We found that the participants (n = 24) performed significantly better in answering the analysis tasks correctly in the model-supported condition (41\% increase in analysis correctness).
Further, participants who reported using the provided traceability information performed better in giving evidence for their answers (315\% increase in correctness of evidence).
Finally, we identified three open challenges of using DFDs for security analysis based on the insights gained in the experiment.

\end{abstract}



\section{Introduction}

\noindent
Dataflow Diagrams (DFDs) are integral to many software security analysis techniques.
For instance, they are required by prominent security assessment techniques~\cite{Sion18, Hernan06, Microsoft16, Torr05}.
They are also the program representation chosen in many model-based security approaches~\cite{Abi-Antoun07, Abi-Antoun10, Berger16, Cao24_catma, Tuma19, Chen17, Stojanovic20, Li19c}. 
As such, they can help software engineers build more secure software systems.
However, their impact on security analysis by their mere provision has not been investigated before to the best of our knowledge. 
DFDs offer a high-level yet detailed representation of applications' architecture. 
Enriched with annotations representing, e.g., employed security mechanisms, they offer easy accessibility of the architectural security.
We hypothesize that providing users with a DFD of an application enables them to analyse the application's security properties with higher correctness.

To investigate this hypothesis, we conducted an empirical experiment.
This paper reports on our findings.
The experiment was performed with master students who solved tasks related to software security analysis activities. 
For this, they received the source code and a textual description of an open-source microservice application, and six tasks to answer.
We chose microservice applications as the target of analysis because the distributed nature of this architectural style poses additional challenges in terms of cognitive load to security analysts.
Systems following the microservice architecture split their codebase into multiple independent microservices, where each fulfils a part of the business functionality and can be independently developed and deployed~\cite{Dragoni16_microservices_yesterday_today_tomorrow, Lewis14_fowler_microservices}.
The resulting codebase can be challenging to oversee in analysis scenarios.

The experiment followed a within-groups design.
In the model-supported condition, participants received a DFD and traceability information of the analysed application in addition to the source code and textual description provided in the control condition.
We chose DFDs created by \textit{Code2DFD}, a tool for the automatic extraction of DFDs from source code~\cite{Schneider23code2dfd}, as these contain extensive security annotations.
We infer insights on the impact of DFDs on security analysis by comparing the participants' performance in analysis correctness, correctness of evidence, and time in the two conditions.
Specifically, we answer the following research questions:

\question{\hspace{-0.5mm}\textbf{1}: Do security-annotated architectural models of applications, specifically DFDs, support developers in the security analysis of the applications?}

\noindent A crucial part of security analysis is identifying and localizing security (and other) features in source code.
To assess whether models with extensive security annotations can support developers and security experts in this activity, the participants in our experiment solved tasks that require the identification of implemented security mechanisms and other relevant system properties.
We quantified their answers and compared the scores between the two conditions.
Additionally, we asked them about the perceived usefulness and analysed the answers.

\question{\hspace{-0.5mm}\textbf{2}: Does access to and use of traceability information improve the ability of the security analysts to provide correct evidence for their analysis findings?}

\noindent Traceability information establishes the validity of model items by referencing corresponding artefacts in the source code.
It can provide value for security analysis, since in scenarios such as software security assessment or certification, evidence has to be given for the reached findings.
The participants provided evidence for their answers in the form of locations in the source code.
We examined its correctness and compared the scores of those participants who reported using the traceability information frequently and those who did not.

\question{\hspace{-0.5mm}\textbf{3}: What is the experience in using security-annotated DFDs for security analysis, specifically concerning the usefulness and accessibility?}

\noindent Users' acceptance of offered tools and techniques is crucial for using resources such as DFDs for security analysis.
Thus, the participants' perceived usefulness of the DFDs is essential to judge their suitability for real-world application.
Further, the information presented by DFDs has to be conveyed to the users efficiently and accessible.
To judge this aspect, we asked the participants about their experience using the DFDs in the model-supported condition and analysed their responses.

\question{\hspace{-0.5mm}\textbf{4}: What are the open challenges of using security-annotated DFDs in the context of security analysis?}

\noindent Based on the insights gained during the empirical experiment and the analysis of the results, we identified and formulated open challenges that should be addressed in future work.

The rest of this paper is structured as follows: Section~\ref{sec:background} introduces the used DFDs; Section~\ref{sec:study_design} describes the experiment's design, i.e., the methodology; the results are presented in Section~\ref{sec:results} and discussed in Section~\ref{sec:discussion}; Section~\ref{sec:limitations} describes limitations of this work; Section~\ref{sec:related_work} presents related work; and Section~\ref{sec:conclusion} concludes the paper.


\section{Characteristics of the Used DFDs}
\label{sec:background}

\begin{figure}
    \centering
    \includegraphics[width = 0.9\linewidth]{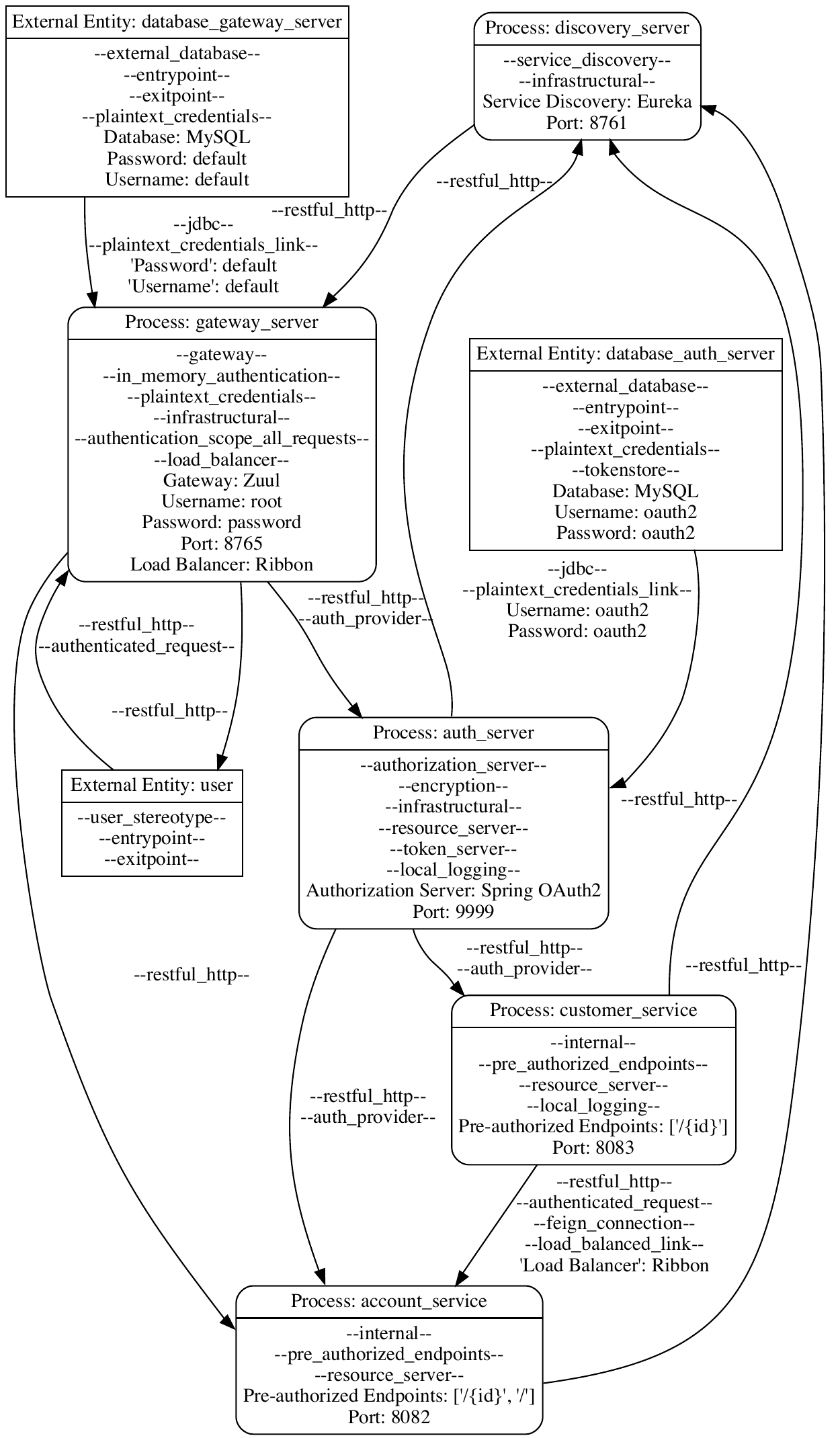}
    \vspace{-2mm}
    \caption{Example DFD provided to participants in model-supported condition.}
    \vspace{-3mm}
    \label{fig:example_DFD}
\end{figure}

\begin{figure}[!ht]
    \centering
    \includegraphics[width = 0.85\linewidth]{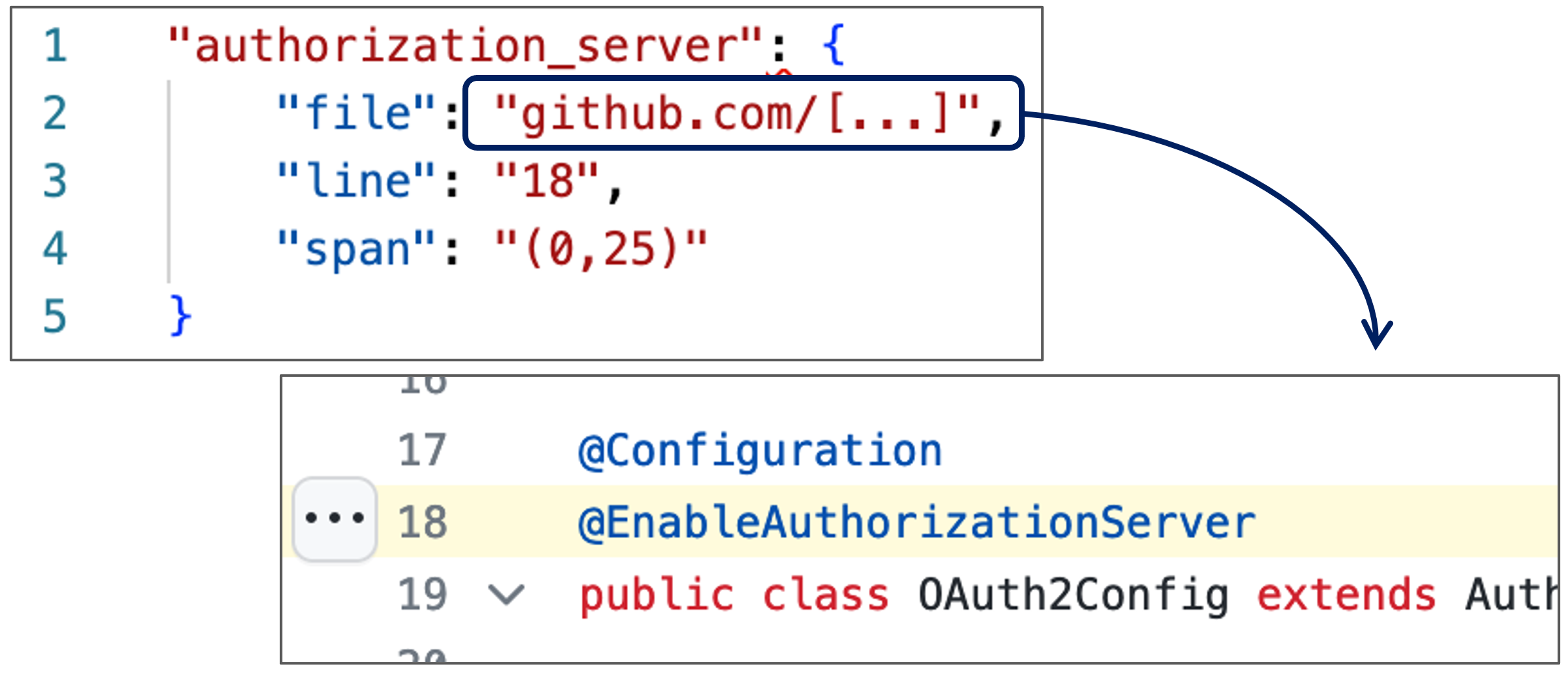}
    \vspace{-2mm}
    \caption{Screenshots of an example traceability information for the annotation \textit{authorization\_server} (top) and the target of the URL (bottom).}   
    \vspace{-5mm}
    \label{fig:traceability_example}
\end{figure}

\noindent 
Since no standard specification for DFDs exists, the various styles of DFDs found in the model-based literature differ in their characteristics.
Here, \textit{style} refers to the types of model items, their presentation and richness of detail, and the scope of considered system components.
All DFD styles share the four base item groups \textit{external entities}, \textit{data flows}, \textit{processes}, and \textit{data stores}~\cite{DeMarco79}.
Many approaches include additional model items to increase the models' expressivity~\cite{Tuma19, Sion18, Tuma18}.
In our experiment, we used DFDs from a dataset published by Schneider et al.~\cite{Schneider23microsecend}. 
An example is shown in Figure~\ref{fig:example_DFD}.
The style is the same as those generated by an automatic extraction approach by Schneider and Scandariato~\cite{Schneider23code2dfd}.
Two of the DFDs' properties stand out compared to other styles, the included \textit{annotations} and the \textit{traceability information} for model items.
Annotations in the DFDs provide information about the corresponding system's security and other properties.
The annotations represent implemented security mechanisms (e.g., encryption or authorization mechanisms), deployment information (e.g., ports or addresses), or other system properties (e.g., used technologies and frameworks). 
Annotations are associated with model items from the four base item groups mentioned above.
That means that base model items (such as, e.g., a service) are augmented with the annotations.
Figure~\ref{fig:example_DFD} shows examples of this (see the \textit{-{}-annotation-{}-} and \textit{Key: Value} annotations).
Traceability information links model items to source code by pointing to code snippets that prove the existence of the model item. 
Figure~\ref{fig:traceability_example} shows as an example the traceability information for the annotation \textit{authorization\_server} as well as a screenshot of the target of the contained URL, i.e., the line of code on GitHub.

\section{Study Design}
\label{sec:study_design}

\noindent
We designed and conducted an empirical experiment to answer the formulated research questions.
We consulted established sources of guidelines for empirical research in the design of the experiment (\cite{Kitchenham02_guidelines_empirical, Juristo01_experimentation, Wohlin12_experimentation}).
All materials as well as the results are available in the replication package~\cite{replication_package}.

\subsection{Setup}
\noindent 
The experiment followed a within-groups design, where participants participated in both conditions.
The study was performed in two 90-minutes lab sessions of a master's course at a university in subsequent weeks. 
The participants were randomly assigned to one of the two groups, G1 and G2.
A different application was given as the target of analysis in the two weeks to mitigate learning effects, \textit{App 1} in week 1 and \textit{App 2} in week 2.
The two sessions were structured as follows:
\vspace{-2mm}

\begin{table}[!ht]
\centering
    \label{tbl:study_setup}
    \begin{tabular}{c|c|c|} 
    
    & \textbf{Week 1} & \textbf{Week 2} \\ 
    \cline{2-3}
    & App 1 & App 2 \\
    \hline
    \textbf{G1} & control condition & model-supported condition \\
    \hline
    \textbf{G2} & model-supported condition & control condition \\
    \hline
    \end{tabular}
\vspace{-3mm}
\end{table}

\noindent
In the \textit{model-supported condition}, the participants performed tasks with access to a DFD and corresponding traceability information, whereas in the \textit{control condition}, they performed a similar set of tasks without this additional support.
They were supervised during the sessions and were discouraged from talking to each other about the experiment.
Google Forms surveys were used to provide the tasks and gather the answers.

\subsection{Tasks}

\begin{table*}
\centering
    \caption{Tasks for App 2. Service names and number of connections slightly differ for App 1. \vspace{-2mm}}
    \label{tbl:tasks}
    \begin{tabular}{c p{13.5cm}} 
    \toprule
    \textbf{ID} & \textbf{Task Description} \\ 
    \midrule
    1 & What is the port number of the microservice \textit{gateway\_server}? \\
    2 & What library is used to implement the API Gateway (service \textit{gateway\_server})? \\
    3 & There are two outgoing connections from the service \textit{customer\_service}. To which services are they connected? \\
    4 & Are these outgoing connections encrypted, i.e., sent with HTTPS? \\
    5 &  Do all business logic services check whether incoming requests are authorized / authenticated?  \\
    6 &  Which service handles the authorization?  \\
    \bottomrule
    \end{tabular}
\vspace{-4mm}
\end{table*}

\noindent
Table \ref{tbl:tasks} lists the analysis tasks given to the participants.
They were chosen such that they resemble common security analysis activities.
The first two tasks are not specific to security but are relevant nevertheless since they foster a required comprehension of the analysed system. 
The tasks cover three different kinds of questions, the participants had to find:

\begin{itemize}
    \item general information about single services (tasks 1~\&~2)
    \item information about security mechanisms of single services (tasks 3 \& 4)
    \item information about system-wide security mechanisms (tasks 5 \& 6)
\end{itemize}

\noindent 
For all tasks, the participants were asked to provide evidence for their answers via a reference to the code.

After completion of the technical tasks, the participants were also posed an open question about their experience with using the resources they had been provided.
In the first week, they were further asked questions concerning their expertise.

As the target of evaluation for the security analysis, we chose two open-source applications from a list published by Schneider et al.~\cite{Schneider23microsecend}.
The applications are referred to as App1~\footnote{github.com/anilallewar/microservices-basics-spring-boot} and App2~\footnote{github.com/piomin/sample-spring-oauth2-microservices/tree/with\_database}. 
These applications were selected based on two properties: (i) high architectural similarity between the two in order to enable an accurate comparison of participants' performance between the two sessions, and (ii) sufficient architectural complexity to allow insightful and relevant tasks.
The two applications exhibit a high degree of similarity concerning their architecture, size, and used technologies.
They incorporate some of the most prevalent microservice patterns and employ widely adopted technology solutions for Java-based microservice development.
For instance, they both realize an API Gateway with Zuul, authentication with OAuth 2.0, a load balancer with Ribbon, and service discovery with Eureka.
Consequently, the applications are a suitable representation of open-source microservice applications developed in Java.

\subsection{Provided Application Artefacts}

\noindent
All participants were provided the source code and a basic textual description of the analysed application.
The textual description is an explanatory document that was created based on the code and information provided by the developers of the applications.
To mitigate a potential influence on the experiment's outcome based on different qualities of the textual descriptions of the two applications, they were created in identical structure and contain the same basic information about the corresponding application.
They illustrate the basic architectural design of the applications.
We remark that some tasks could be answered based on these documents.
The code was provided via GitHub.
Specifically, the applications' repositories were forked to remove the original documentation, which could otherwise influence the outcome.

In addition to the code and textual description, the participants in the model-supported condition received the DFD and the traceability information of the application to be analysed. 

\subsection{Participants}
\label{sub:participants}
\noindent
The experiment's participants comprised 24 students enrolled in a master's level software security course at \textit{Hamburg University of Technology}. 
They originate from various disciplines, all incorporating computer science to a large degree.
The students were informed about the empirical experiment two months in advance and were invited to participate.
\begin{figure*}
    \centering
    \subfigure[]{\includegraphics[width=0.32\textwidth]{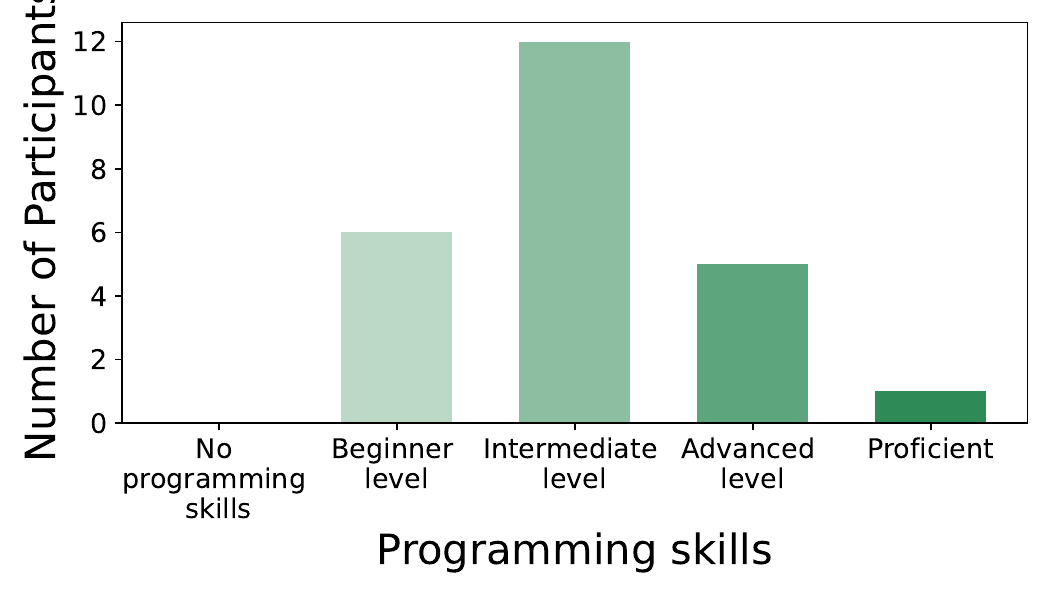}} 
    \subfigure[]{\includegraphics[width=0.32\textwidth]{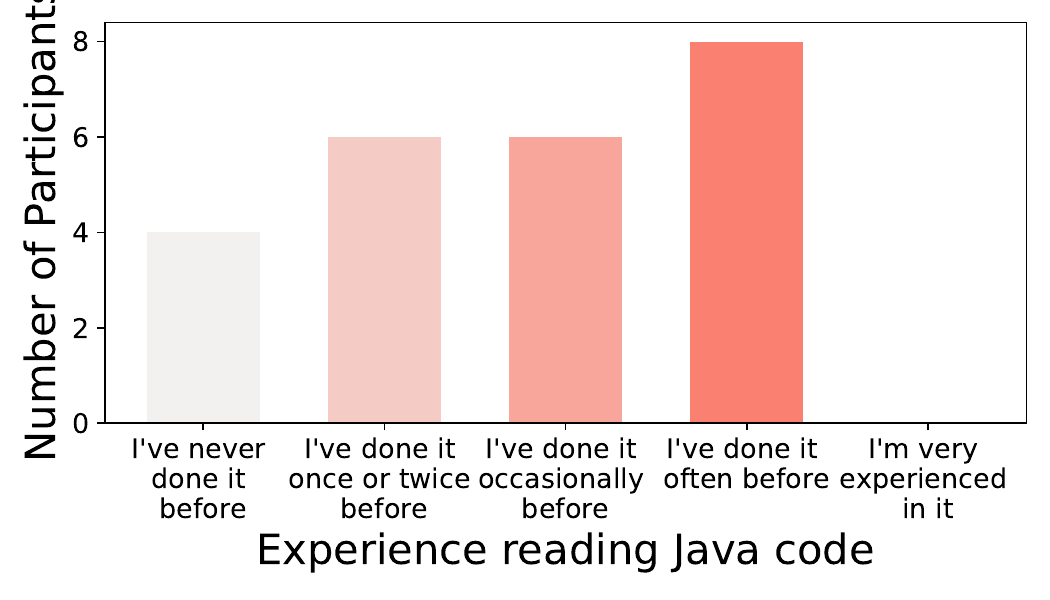}} 
    \subfigure[]{\includegraphics[width=0.32\textwidth]{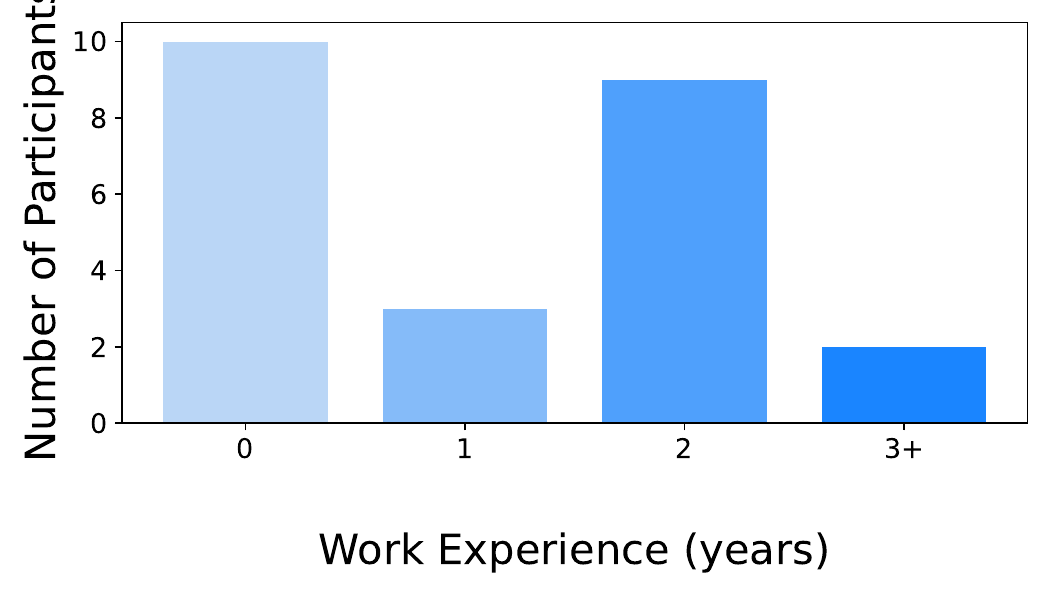}}
    \vspace{-2mm}
    \caption{Participants' (a) programming skills, (b) experience in reading Java code, and (c) work experience as developers. All self-reported.}
    \label{fig:participant_expertise}
    \vspace{-5mm}
\end{figure*}
Figure~\ref{fig:participant_expertise} shows the participants' programming knowledge (a), experience with reading Java code (b), and work experience (c). 
This information was reported in a self-assessment at the end of the first week's session. 
Based on the results (“intermediate level” being the answer most often given to the question of programming skills; little experience with reading Java code; and an average work experience of 1.1 years), we deduced that the participants were on average advanced beginners in software development and had moderate experience in analysing Java code. 
Accordingly, they represent well the target population of the experiment.
The goal of the experiment was to investigate the effect of architectural models on the performance of users with low expertise in software security analysis, for example novice developers.
The metrics in Figure~\ref{fig:participant_expertise} fit well to such users.
Furthermore, using students as proxies for the target population is a common practice in empirical software engineering and has been shown to be a suitable method~\cite{Kitchenham02_guidelines_empirical, Salman15_students_representative, Svahnberg08_student_subjects, Falessi18_students_in_experiments}.

\subsubsection{Incentives}
The students were generally incentivized to participate in the course's lab sessions, independent of the empirical experiment. 
For this, they were rewarded with a 5\% bonus on their final exam, granted upon attending all of the lab sessions in the semester except for one (a common practice at the university). 
Further, they were encouraged to participate because the sessions were relevant for the final exam, and the participants could hone the required skills there.
The lab sessions where the empirical experiment was conducted were akin to all other lab sessions in these regards.
Consequently, the experiment's impact on students' grades was consistent with that of the other lab sessions in this course.
No other incentives were pledged or given.

\subsubsection{Preparation}
To prepare the participants, a 90-minute lecture before the lab sessions was dedicated to introducing them to the topic (available in this paper's replication package~\cite{replication_package}).
The lecture covered key concepts relevant to the experiment.
The primary focus was on the origin of software vulnerabilities and methods for detecting them. 
The lecture also encompassed topics such as DFDs, microservice architectures, and security considerations in microservice applications. 
Following this lecture, the students were expected to possess the required knowledge to undertake the experiment.
Their attendance at the lecture was recorded and was a prerequisite for participating in the experiment.

\subsubsection{Informed consent and ethical assessment}
All participants read and signed an informed consent form before the experiment, informing them that they are the subjects of an empirical experiment, that they participate voluntarily, that they do not have to expect any negative consequences whatsoever if they do not participate, and that they can retract their consent at any time.
To ensure the experiment's ethical innocuity, it was assessed by the \textit{German Association for Experimental Economic Research e.V.} before execution. 
A description of the planned experiment and its design was approved under grant nr. \emph{2pxo1bap}. 
The certificate can be accessed via \url{https://gfew.de/ethik/2pxo1bap}.

\subsection{Measurement}
\noindent 
To evaluate the participants' performance, three metrics were introduced.
The \textit{analysis correctness} represents the ability to provide correct answers to the tasks.
The \textit{correctness of evidence} measures whether the evidence that the participants provided as support for their answers points to a code snippet that justifies their answer.
Both are numerical scores derived from the participants' responses.
Additionally, we measured the time spent on solving the tasks.
The three metrics (analysis correctness, correctness of evidence, and time) were calculated for each participant in both conditions separately.

\subsubsection{Analysis Correctness}
We quantified the given answers concerning the analysis correctness by manually checking the participants' responses.
To remove subjectivity, we created a reference solution that was used to check the answers.
It was created prior to the execution of the experiment.
The DFDs and source code of the applications have been analysed to create the reference answers, which were afterwards confirmed by conducting technical documentation of the code libraries used in the applications, information provided by the developers of the applications, and other typical online resources.
This process was performed by the first author and validated afterwards by two additional authors.
After the experiment, the participants' responses were mapped via the reference solution to a table indicating correct and incorrect answers.
Each response was examined manually, compared against the reference solution, and true answers were marked in the table.
We further reviewed answers that did not match the reference solution to check whether they were correct.
For this, various typical online resources were conducted to verify whether the specific answer applies to the task.
Each correctly given answer gives a score of 1. 
There is a peculiarity for some tasks.
Task 3 asks for a list of connections between services, and tasks 4 and 5 ask whether a property applies to each item on a given list.
Consequently, these tasks each required multiple distinct responses.
All responses were checked individually.
Then, to allow a more detailed and nuanced evaluation, we converted the results to scores.
A score of 0 was assigned for no correct responses, a score of 1 was given for partially correct responses (meaning that some but not all responses of a task were correct), and a score of 2 was awarded when all responses were correct.
With three tasks giving a maximum of one point and three tasks giving a maximum of two points, the overall highest achievable score in analysis correctness is~9.

\subsubsection{Correctness of Evidence}
The traceability information that is contained in the used DFDs constitutes a reference solution for quantifying the correctness of the evidence given by the participants.
Each given evidence was checked manually for matches to this reference solution.
Here, we employed some tolerance in accepting evidence as correct. 
For example, when participants referred to a block of code slightly larger than the lines of code needed to prove an answer, we still accepted this as correct (e.g., referring to a method consisting of some lines of code instead of referring to a single line of code in that method).
We carried out a further validation check, similar to the quantification of the analysis correctness.
Each provided evidence that differed from the reference solution was checked manually whether it supported the given answer or not.
The first author carried out the above steps.
As for the analysis correctness, each correct evidence gives a score of 1.
Again, for tasks 3, 4, and 5, the multiple distinct responses were converted into a score of 0, 1, or 2 for no correct, partially correct, and all correct responses, respectively.

\subsubsection{Time}
To measure the time spent on solving the tasks, the participants were asked to record the current time when starting and finishing to work on the tasks.
We calculated the time metric based on these answers (i.e., the period of time between the start and finish of solving the tasks).

\subsubsection{Reported Usefulness}
The DFDs' usefulness as reported by the participants was assessed via the open question about the participants' experience in using the DFDs that was posed after the technical tasks. 
We qualitatively analysed all answers' general intents (positive/negative feedback) and identified recurring topics manually.
This analysis was performed by the first author and verified by two further authors.

\subsection{Statistical Tests}
\label{sub:statistical_tests}
\noindent
Throughout the analysis, the difference of scores between two groups was checked for statistical significance with a Wilcoxon-Mann-Whitney test.
Before, a Shapiro-Wilke test was used to verify that the data does not follow a normal distribution.
Hence, no parametric tests could be used.
The assumptions for the Wilcoxon-Mann-Whitney test (the samples are independent, random, and continuous and the sample size is sufficiently large) are met in our experiment.


\section{Results}
\label{sec:results}

\subsection{Analysis Correctness}

\begin{figure}
    \centering
    \includegraphics[width = \linewidth]{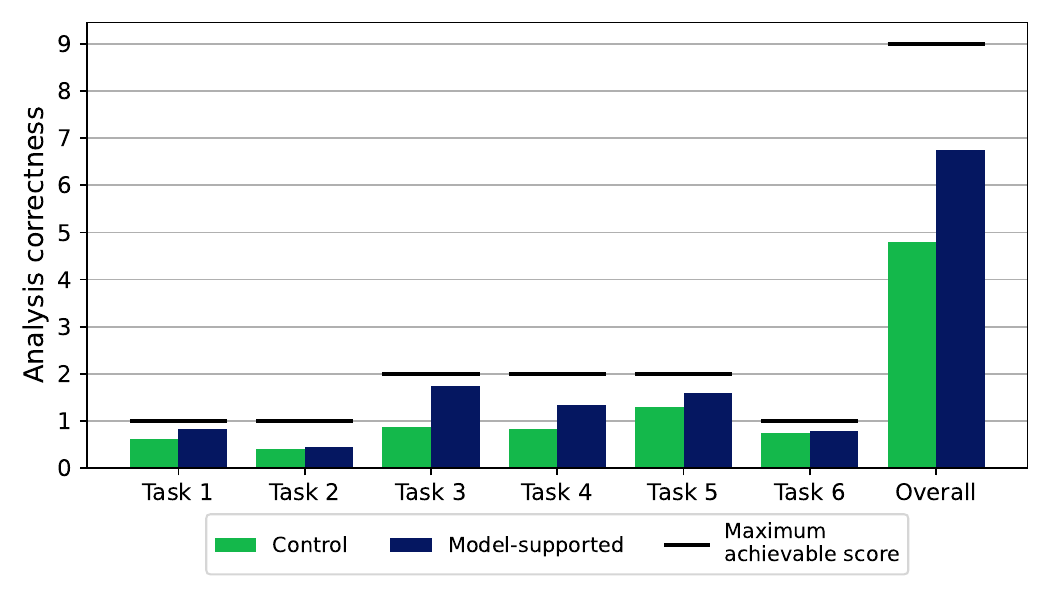}
    \vspace{-8mm}
    \caption{Comparison across the two treatments of the participants' average score in analysis correctness. Per task and overall.}
    \label{fig:average_score}
    \vspace{-4mm}
\end{figure}

\noindent
Figure~\ref{fig:average_score} presents the average score in analysis correctness that the participants achieved in the two conditions. 
The figure shows that the participants performed better in the model-supported condition, both overall and in every individual task.
For all tasks, the average score is 6.75 out of a possible 9 in the model-supported condition compared to 4.79 in the control condition, a 41\% higher average.
The applied statistical test (compare Section~\ref{sub:statistical_tests}) indicates a statistically significant difference between the two conditions' average scores in analysis correctness overall (p = 0.0025).
These results provide the following answer to RQ1:

\answer{\hspace{-0.5mm}\textbf{1}: In the context of our experiment, providing a security-annotated DFD of the system to be analysed improved participants' analysis correctness in solving security analysis tasks. 
We observed a statistically significant (p~=~0.0025) improvement of 41\% on average.}

\noindent For some individual tasks, the difference in the average scores is only marginal (task 2: 0.42 vs. 0.46 $\estimates$ 10\% improvement in model-supported condition; task 6: 0.75 vs. 0.79 $\estimates$ 5.6\% improvement).
In Section~\ref{sec:discussion}, we discuss whether the nature of the tasks might be an indication of the extent to which a DFD improves the score in analysis correctness.
However, even though the improvement is not statistically significant for all individual tasks (statistical significance only for task 3 with a p-value of 0.0003), Figure~\ref{fig:average_score} clearly shows a trend of improved performance in the model-supported condition.

\subsection{Correctness of Evidence}

\begin{figure}
    \centering    
    \includegraphics[width = \linewidth]{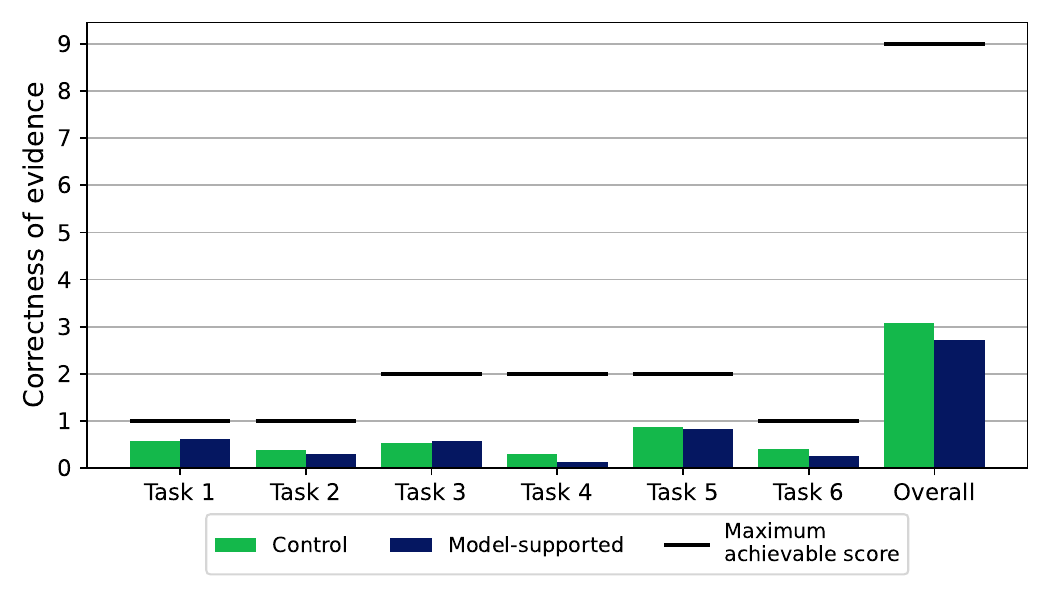}
    \vspace{-8mm}
    \caption{Comparison across the two treatments of the participants' average score in correctness of evidence. Per task and overall.}
    
    \label{fig:average_evidence}
    \vspace{-3mm}
\end{figure}

\noindent
Figure~\ref{fig:average_evidence} presents the average score in correctness of evidence achieved by the participants. 
There are only small differences in the average scores for the model-supported and control condition.
With an average of 3.08 out of a possible 9 in the control condition compared to an average of 2.71 in the model-supported condition (-12\%), the participants performed better in the control condition, albeit without statistical significance (p = 0.52).
Task 4 has the lowest average correctness of evidence of the individual tasks (average of 0.29 out of 2 in the control condition and 0.13 in the model-supported condition), while task 1 has the highest average (0.58 out of 1 in the control condition and 0.63 in the model-supported condition).

\subsection{Use of Provided Artefacts}

\begin{figure}
    \centering
    \includegraphics[width = \linewidth]{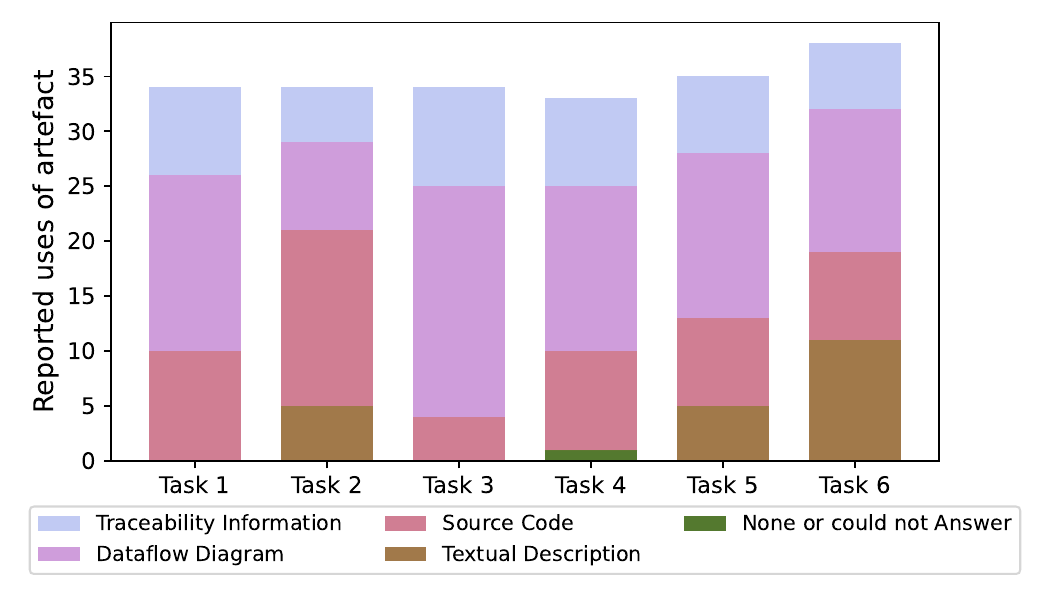}
    \vspace{-8mm}
    \caption{Reported usage of provided artefacts per task (in the model-supported condition, where all artefacts were available).}
    \label{fig:resource_usage}
    \vspace{-4mm}
\end{figure}

\noindent 
After each task, the participants were asked to name all provided artefacts that they used to answer the task.
Figure~\ref{fig:resource_usage} shows the reported usage per task in the model-supported condition.
We focus on this condition because, there, the participants had access to all artefacts.
Note that the participants had the possibility to name multiple artefacts per task.
Overall, the DFD was used the most, with a total of 88 answers. 
The source code was used 55 times, the traceability information 43 times, and the textual description 21 times.
The numbers show, that the participants did not solely rely on the provided DFD, but instead also referred to the source code in many cases and the textual descriptions in some cases.
This could indicate, that the participants verified information they found in the DFD by checking the corresponding part of the code.

\subsection{Influence of Use of Artefacts on Scores}

\noindent
We investigated whether the participants' usage of the provided artefacts had an influence on their performance.
Although they were provided more artefacts in the model-supported condition, this does not necessarily mean that they used them all. 
The answers to the tasks could be found with more than one of the artefacts.
Thus, the influence of single artefacts on the performance is not necessarily reflected in the comparison of outcomes between the two conditions. 
For example, participants in the model-supported condition could have not used the provided DFD to answer the tasks.
Consequently, we compared the average scores in analysis correctness and in the correctness of evidence between two groups of participants for each artefact.
To the group \textit{Using Artefact} we assigned all participants that reported using the artefact in more than 50\% of the tasks (4 or more).
The group \textit{Not Using Artefact} contains those participants who reported using it less (3 or fewer).
We considered only the outcomes from the participants in the model-supported condition, since only here they had access to all artefacts.
The grouping and analysis were done separately for each artefact, thus, the cardinality and members of the groups differ between artefacts. 

\begin{figure}
    \centering
    \includegraphics[width = \linewidth]{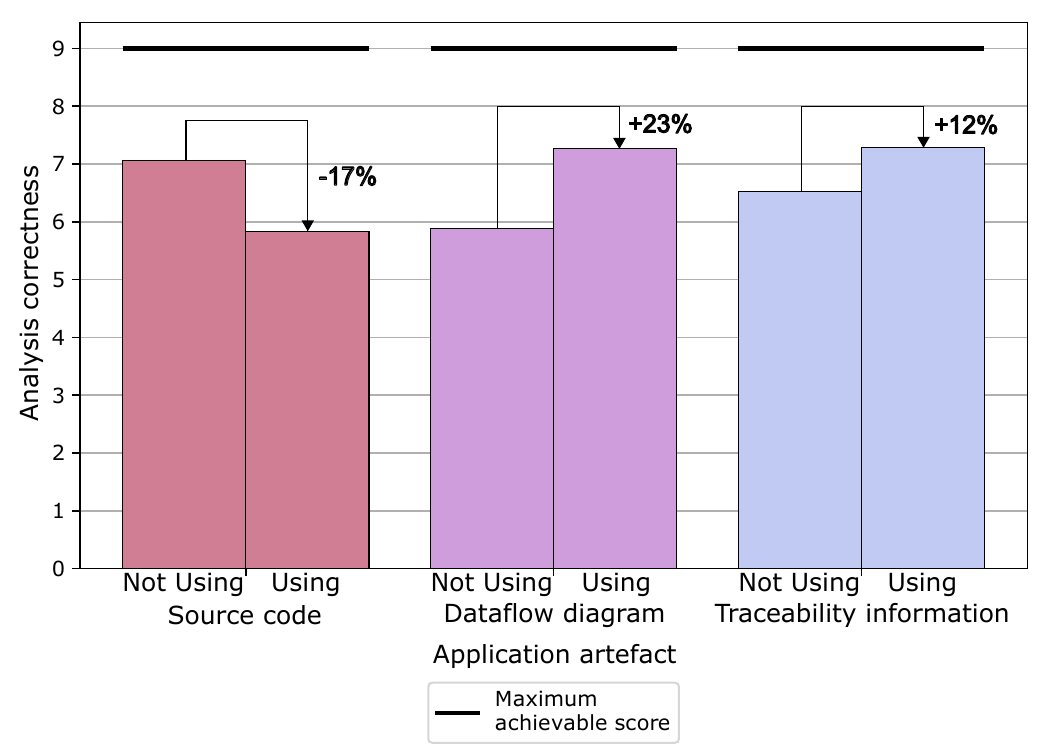}
    \vspace{-7mm}
    \caption{Average scores in analysis correctness of those participants that reported using an artefact in more than 50\% of the tasks (Using Artefact) and those that reported using it less (Not Using Artefact).}
    \label{fig:average_scores_resources}
    \vspace{-6mm}
\end{figure}

Figure~\ref{fig:average_scores_resources} presents the average scores in analysis correctness of the two groups per application artefact.
The figure shows that the Using Artefact group performed better compared to the Not Using artefact group for the artefacts DFD (+23\% in score) and traceability information (+12\%), while they performed worse for the source code (-17\%).
\begin{figure}
    \centering
    \includegraphics[width = \linewidth]{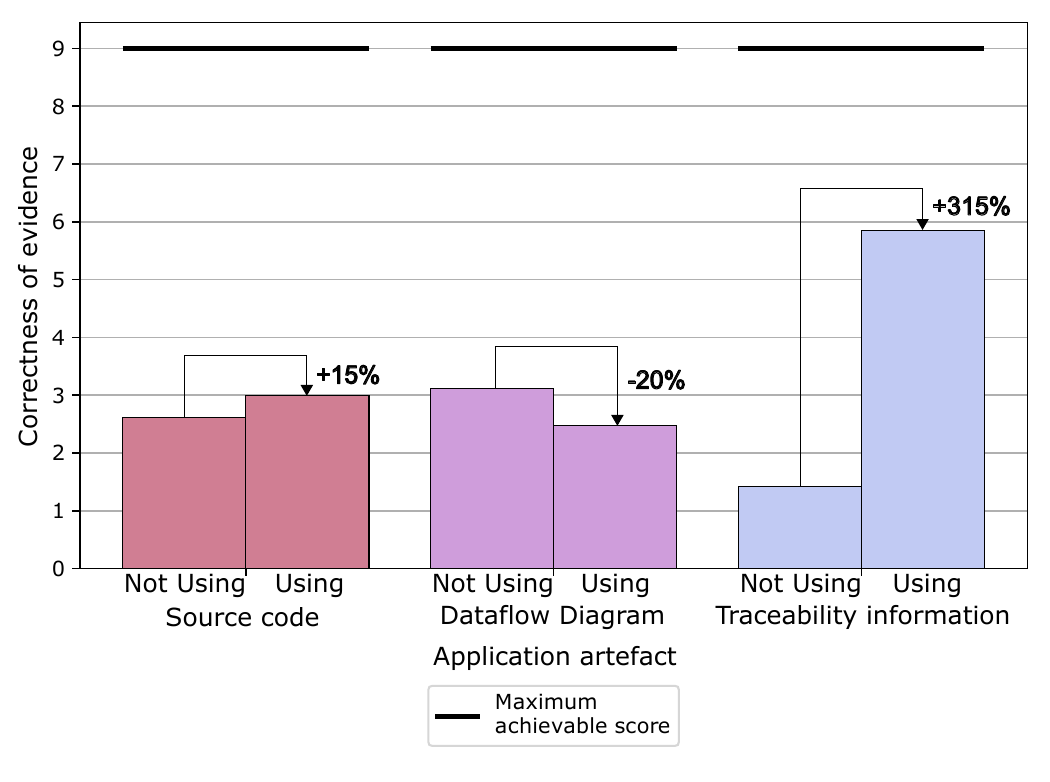}
    \vspace{-9mm}
    \caption{Average scores in the correctness of evidence of those participants that reported using an artefact in more than 50\% of the tasks (Using Artefact) and those that reported using it less (Not Using Artefact).}
    \label{fig:average_evidence_resources}
    \vspace{-7mm}
\end{figure}
Figure~\ref{fig:average_evidence_resources} presents the results of this analysis for the correctness of evidence.
For the artefact source code, the Using Artefact group achieved a 15\% higher score in correctness of evidence than the Not Using Artefact group.
For the use of DFD, they performed worse than the Not Using Artefact group (-20\%).
The highest difference, however, is seen in the traceability information.
The Using Artefact group achieved a 315\% higher average score in correctness of evidence than the Not Using Artefact group.
We answer RQ2 based on these results since they distinguish between the use of the DFD and traceability information.
In the results above, this distinction could not be made because the traceability information is an integral part of the DFDs and its isolated impact on the analysis could not be measured.

\answer{\hspace{-0.5mm}\textbf{2}: Using traceability information significantly improved the correctness of evidence given for answers.
On average, participants that used this artefact in more than half of the tasks achieved a 315\% higher correctness of evidence compared to participants that used it less than that.}

\subsection{Time}
\noindent
All participants were able to complete the tasks in the allotted 90 minutes.
Their average time to complete all tasks was 34 minutes in the control condition and 35 minutes in the model-supported condition.
No notable difference was observed. 
To examine a possible correlation between performance and time spent to finish the tasks, we also created a scatter plot visualizing their scores against their time. 
No correlation between scores and time could be visually identified.

\subsection{Perceived Usefulness and Usability of DFDs}
\noindent
The answers given to the open question at the end of the analysis sessions provide insights into the participants' perceived usefulness of the DFDs.
The question asked about positive or negative observations during the experiment.
For participants in the model-supported condition, it explicitly mentioned the usefulness of the DFDs and traceability information.

Out of 23 answers given by participants in the model-supported condition, two were negative, stating that thorough documentation would be preferred and that the DFD was “\textit{a little bit hard to understand at first}”.
Three answers listed both positive and negative experiences, where the negative points were two mentions that finding implementation details was hard (both participants reported using the traceability only in one task) and one that the participant lacked domain knowledge.
A further 14 answers were predominantly positive.
\begin{quote}
    \textit{“Dataflow Diagrams were incredibly helpful, and all questions were answered almost completely from it.”}
\end{quote}

\noindent 
Of the 23 answers, 9 mentioned specific beneficial scenarios for the use of DFDs.
The ability to provide an overview of the system was mentioned 8 times, the benefit of referring to the important places in source code and use of the models as interface to the code was mentioned 3 times, and the reduction of the required domain knowledge was mentioned once.

Mild critique about the accessibility of the DFDs or traceability information was raised in 4 responses, for example:
\begin{quote}
    \textit{“[...] the transfer from the DFD to the traceability information could be made easier by clickable links in the DFD [...]”}
\end{quote}

In summary, the statements made by participants in the model-supported condition include descriptions of the general usefulness of the DFDs, of benefits in finding implementation details via the model items and traceability information, and of their usefulness for architectural details and providing an overview.
The positive feedback outweighed the few negative comments.
Most participants reported the DFDs to be of help in the analysis and to be accessible to use.

Of the answers given after the control condition, only one was positive, stating that the textual description was helpful.
Four others referred to the DFDs (these answers were given in the second week, the participants had thus already performed the session with the DFD), stating that, in comparison, the lack of a DFD was an obstacle during the analysis.
Specifically, they raised concerns about the correctness of their given answers and stated that finding the required features directly in source code was challenging.
\begin{quote}
    \textit{“The traceability file and the DFD were a big help last time, this time I wasn't really sure if I even answered correctly and didn't really know if the evidence I gave was correct. [...]”}
\end{quote}
Six further answers of participants in the control condition in the first week (and thus without the comparison to the model-supported condition) reported negatively about their experience in the experiment.
Specifically, they mentioned a lack of expertise, uncertainty about the given answers, and general difficulties in answering the tasks.
Interestingly, two participants criticized the lack of a “\textit{CFG}” or “\textit{some kind of map of the architecture}”.
This could have been sparked by the introductory lecture where DFDs were addressed but is still seen as an interesting comment.
The obstacles reported by the participants in the control condition give further weight to the positive feedback of those in the model-supported condition.

Based on a qualitative analysis of the participants' statements, we can cautiously judge the perceived usefulness and accessibility of the DFDs to answer RQ3: 

\answer{\hspace{-0.5mm}\textbf{3}: In our experiment, the perceived usefulness and accessibility reported by the participants varied from very positive feedback to mild critiques reporting some confusion.
Overall, the statements focussed on usefulness and were predominantly positive.}
\vspace{-2mm}

\subsection{Open Challenges of DFDs}
\noindent
The above observations of the quantitative and qualitative results allowed us to distill a number of open challenges of DFDs, i.e., current obstacles that would increase the DFDs' positive impact further if solved.
Although these challenges were not explicitly investigated as independent variables in our experiment, they became evident from the results of the experiment, explicit answers given by participants, and observations made during the analysis of the tasks.

\noindent \textit{\textbf{Open Challenge 1: Understandability of Models}}
The participants in our experiment performed with statistical significance better in the model-supported condition and they reported a generally good accessibility.
Nevertheless, concerns were raised about the understandability of the models.
Some participants commented, that they did not understand the model initially or that they did not know what some annotations mean.
A more usable model representation of software systems should consider the accessibility for human users, especially those with lower domain knowledge.

\noindent \textit{\textbf{Open Challenge 2: Presenting Missing Features}}
The DFDs in their current form do not support the explicit presentation of the absence of features or properties.
In the context of security analysis, these could be security mechanisms that are not implemented by a given application.
To enable more comprehensive analysis and increase users' trust, it is important to show that such mechanisms were investigated and are not implemented in the analysed application.
In this context, the challenges are to prove the absence, to decide what features to consider, and how to convey this information to the user.
We see this open challenge as the hardest one to solve, both conceptually and practically.

\noindent \textit{\textbf{Open Challenge 3: Accessibility of Traceability Information}}
The quantitative results of our experiment show, that the traceability information has a positive impact on the correctness of evidence provided for answers to the tasks.
While this is an expected observation, multiple participants also mentioned the usefulness of the traceability information for navigating the source code.
However, it was also mentioned in some answers that the connection to the source code was difficult to follow.
Also, the traceability information was not used by everyone even when it was provided.
We conclude that the ease of use can be improved and that navigating the links to source code should be simplified.
This challenge is of more practical nature and can likely be solved with some clever engineering.

\answer{\hspace{-0.5mm}\textbf{4}: We identified three open challenges of DFDs (understandability of models, presenting missing features, and accessibility of traceability information). If any of these are solved, the positive impact of DFDs on security analysis can be expected to further increase.}

\section{Discussion}
\label{sec:discussion}

\noindent 
At the heart of the conducted experiment lay the question of the impact of providing DFDs and traceability information on the participants' performance.
The results presented in Section~\ref{sec:results} indicate an overall positive impact on the analysis correctness.
The scores improved with statistical significance in the model-supported condition.
Figure~\ref{fig:average_scores_resources} emphasizes this finding.
Participants in the model-supported condition who reported using the provided DFD in more than half of the tasks had a 23\% higher score on average than those who reported using it less. 
A 12\% higher average score for participants using the traceability information is further proof of the usefulness of the DFDs, since the traceability information is one of their core features.
The observed 17\% lower score in analysis correctness for participants who reported using the source code in more than half of the tasks was an unexpected outcome at first sight.
A closer look at the usage of the source code as artefact revealed, that out of 55 responses that mentioned using the source code, 34 (this corresponds to 62\%) did not use the DFD or traceability information in conjunction.
In other words, the source code was predominantly not used alongside the models, but instead as the only artefact to answer a task. 
Consequently, in our experiment, many participants who reported using the source code could also be described as not using the provided models.
With this re-phrasing, the results are another indication of the models' positive impact.

Looking at the individual tasks, the increase in scores differed between them.  
This suggests the question of which type of tasks the DFDs have the most impact on, and how exactly they impact different types of tasks.
We investigated whether the nature of the tasks could be an explanation for the observations, i.e. whether the type of task can indicate how the score is influenced.
We found that the DFDs impacted the analysis tasks in our experiment in different ways.
They are described in the following.
Please refer to Table~\ref{tbl:tasks} for the tasks.

\textit{Providing an Overview:}
Tasks 1, 2, and 6 have fairly simple answers in comparison to the other tasks.
The answer for task 1 (in which the analysis correctness improved by 33\% in the model-supported condition) could be found at two places in the code, either a deployment file indicating the container's port, or a configuration file indicating the service's port.
Both answers were accepted as correct.
In the DFD, the port is shown as annotation to the corresponding node.
Interestingly, the wrong answers given are one of two options. 
One is the port number of a different microservice, which likely showed up when searching for “port” with GitHub's search function.
The other is the port of a database that is only visible in the code as part of the database's URL.
How this answer was reached by participants is puzzling.
For task 6, the improvement of the average score was the lowest of all tasks (0.75 in the control condition and 0.79 in the model-supported condition; 5.6\% increase).
The task has the overall best average scores, likely, because the authorization service's name (“auth\_server”) hints towards the answer of which of the services handles the authorization.
Task 2 could be answered based on the textual description, on the Java annotation that implements the API gateway in code, or on an annotation in the DFD.
A 10\% improvement in average score in analysis correctness was observed from the control condition (0.42) to the model-supported condition (0.46).
The answers lead us to believe that the question might have been formulated such that participants did not fully understand it.
Many of the wrong answers in both conditions stated the used framework (Spring) instead of the library that was asked for (Zuul).
Further, this task had the lowest reported number of usages of the DFD as well as traceability information (compare Figure~\ref{fig:resource_usage}).

The answers and evidences indicate, that DFDs are helpful in providing an overview and presenting the answers to simple questions such as the port number of a microservice.
Evidently, finding \textit{any} port in the code is a simple task in many systems' codebases, however, the answers suggest that finding the \textit{correct} one can be challenging.
Likely, this is heightened by the complexity that the microservice architecture adds to an application's codebase due to its decoupling.
The answers given by participants in the model-supported condition further emphasize this quality of DFDs to provide an overview of the important system components (compare Section~\ref{sec:results}, where this was the benefit most often mentioned by participants).
Simultaneously, for simple tasks with a fairly easy answer, good coding practice such as choosing descriptive identifiers seems to support the analysts well and there is no pressing need to provide a DFD.
Whether this holds true in the analysis of larger applications should be investigated in future work.
The results of task 2 indicate problems in the DFDs' accessibility.
The presented information seems to not be self-explanatory enough for the participants to answer this task reliably, even when the information is contained in the DFDs.

\noindent\fbox{
    \parbox{0.94\linewidth}{%
        \faArrowRight \hspace{0.5mm} The results indicate, that DFDs serve as a means to “navigate the jungle” that is the application's codebase. They provide an overview of the application's architecture and (security and other) features. At the same time, well-chosen identifiers in code can support the solving of simple analysis tasks and the DFDs add less value in this scenario.
        }
}
\vspace{1mm}

\textit{Reducing Required Domain Knowledge:}
To answer tasks 3 and 5 in the control condition, some domain knowledge was needed to correctly grasp the functionality of the relevant code.
Task 3 required the participants to identify three outgoing connections (for App 1, two for App 2) of a microservice.
One is a direct API call implemented with Spring Boot's \texttt{RestTemplate}, another a registration with a service discovery service, and the third a registration with a tracing server (similar for App 2).
Some domain knowledge about these technologies or Java was required to identify them. 
With the DFD at hand, answer the task came down to identifying the correct node in the diagram and noting the three nodes to which there was an information flow.
To answer task 5 without the DFD, participants had to check whether three services (for App 1, two for App 2) refer to the authorization service in a configuration file under an \texttt{authorization} section.
In the DFD, a connection to the authorization server indicated this. 
Again, knowledge about Spring or Java made it easier to find the correct answers without the support of the DFDs.

Task 3 showed the biggest impact of the models, with a doubled average score in analysis correctness (0.875 in control condition and 1.75 in model-supported condition; 100\% increase).
While this task was more difficult to answer than the others without a DFD and the required domain knowledge, the magnitude of the difference is still substantial.
For task 5, the average score in analysis correctness was 1.29 in the control condition and 1.58 in the model-supported condition, a 23\% increase.
The differences show how the DFDs reduce the domain knowledge required for analysis activities. 
However, we hypothesize, that the participants without the DFD could answer the task simply by identifying the keyword “authorization” in the configuration files without checking if the implementation is correct and behaves in the way that is asked for.
We believe, that this led them to achieve an average score without the DFDs that is still high.
Given the scenario in which they solved the tasks (empirical experiment, where answers are expected), this was likely sufficient evidence for them to answer, independent of whether their domain knowledge was profound enough to fully understand the workings.

\noindent\fbox{
    \parbox{0.94\linewidth}{%
        \faArrowRight \hspace{0.5mm} Our interpretation of the results is that DFDs are especially helpful in scenarios where a lack of domain knowledge about the analysed application's framework, libraries, etc. hinders the identification of features and system components.
        The DFDs' ability to shed light on properties shaded by a curtain of domain knowledge seems to be one of their core virtues.
    }
}

\textit{Indicating Absence of Features:}
Despite the open challenge 2 (presenting missing features in the DFDs, see Section~\ref{sec:results}), the results also indicate that the DFDs in their current form already support users in answering tasks concerning the absence of features in the code.
Task 4 was different from the other ones in that the challenge lay not in finding an artefact in the code but instead the absence of it. 
The task asked for the presence of encryption in two connections (for App 1, three for App 2) between services.
The correct answer to all of them was “No”. 
The average score in analysis correctness was 0.83 in the control condition and 1.33 in the model-supported condition out of a possible score of 2 (60\% increase).
The difficulty in this task also became apparent when looking at the results for the evidence.
The participants achieved an average score in correctness of evidence of 0.042 in both conditions.

\vspace{1mm}
\noindent\fbox{
    \parbox{0.94\linewidth}{%
        \faArrowRight \hspace{0.5mm} Although the DFDs still face the open challenge of presenting missing features, their current form already supports users in answering tasks that require identifying the absence of features in code.
        }
}
\vspace{1mm}

In summary of the discussion of the results, we see that the DFDs had a positive impact on the scores in different types of tasks.
Specifically, they provide an overview of the analysed application, they reduce the required domain knowledge, and they can indicate the absence of features in the application.
The highest increase in scores is seen for tasks where some domain knowledge was needed to answer them without the DFDs.
The only task where the improvement of the analysis correctness in the model-supported condition was neglectable was a simple task where descriptive identifiers in code indicated the answer.


\section{Threats to Validity}
\label{sec:limitations}

\noindent 
\emph{Internal validity:} 
With a large group of university students as participants, collaborations during or between the sessions and resulting cross-contamination cannot be ruled out completely.
As mitigation, we strictly discouraged collaborations and conversations about the study and supervised the analysis sessions.
Learning effects or the possibility of preparing for the tasks were mitigated with the employed within-groups design where the scenarios switched over the two sessions and with the use of different applications.
With 90-minute long sessions, experimental fatigue is limited.
The random assignment to the groups G1 and G2 limits selection bias.
Some of the analysed data (timestamps, experience, resource usage) is self-reported, and we have to rely on its correctness.
The encouragement of positive as well as negative feedback and the often-repeated reassurance of full anonymity of the answers were used to increase the reliability of the data.
By making participation voluntary and using only the standard incentive for attending the lab sessions, it is possible that we have attracted mainly students who show high motivation and are at the top of their class.
This could have had distorting effects on the results and could not be reasonably mitigated.

\emph{External validity:} The conclusions drawn in this paper might not entirely map to other scenarios or populations.
The tasks used as examples of security analysis activities might differ from real-world use cases and thus influence the shown effects.
Further, the experiment focused on microservice applications written in Java.
We chose Java applications because it is the most used programming language for open-source microservice applications.
The analysis of systems that follow a different architectural style or are written in another programming language could show other outcomes.
The number of participants (24) is relatively small.
We chose robust statistical methods that are suitable for the sample size and discussed the impact of the participants' experience and the choice of tasks.
The participants' expertise in security analysis is rather low.
Thus, the effects described in this paper might not be observed for other users, e.g., with more experience.
However, the use of DFDs is not confined to security experts, hence rendering the participants a suited population for the experiment.
Finally, experiments with practitioners instead of students could lead to different results, however, it is a common practice and has been shown to produce valid results as well (see Section~\ref{sub:participants}).

\emph{Construct validity:} We measured the participants' performance in terms of correctness and time, which are common and objective metrics for such experiments.
They relate to the practical use-case of the investigated effects directly.
The analysis correctness is crucial in security analysis to ensure accurate security evaluations and, consequently, secure systems.
The time serves as a measure of productivity and efficiency.
Other constructs were disregarded but could be suited as well.

\emph{Content validity:} The tasks concerned the key security mechanisms implemented in the analysed applications.
These or similar tasks would be part of a real-world security analysis.
However, other tasks might also be important in this context.

\emph{Conclusion validity:} The responses to the tasks were given in free-text fields. 
Although we did not identify such ambiguities in quantifying the responses, it is possible that some answers were phrased in a way that was interpreted incorrectly.
A more restrictive way of collecting the answers could have increased the conclusion's validity.


\section{Related Work}
\label{sec:related_work}

\noindent 
Although DFDs are used for different aspects of security analysis, no related work could be found that investigates their direct impact on the correctness of the analysis.
Publications for other model types exist.
For example, a considerable body of empirical research on Unified Modeling Language (UML) diagrams has been published~\cite{Budgen11_uml_slr}.
A number of experiments have been conducted to investigate whether users' comprehension of the modelled systems increases with UML diagrams.
Gravino et al.~\cite{Gravino15_code_comprehension_uml, Gravino10_empirical_investigation_code_comprehension} observed a positive impact of the models, while experiments by Scanniello et al.~\cite{Scanniello15_comprehensibility_family} did not show such an improvement (the authors attribute this to the type of UML diagrams, which had little connection to the code since they had been created in the initial requirements elicitation phase of the development process).
In an experiment by Arisholm et al.~\cite{Arisholm06_impact_uml}, code changes performed by participants with access to UML documentations showed significantly improved functional correctness.
Other researchers investigated the impact of specific properties of UML diagrams on users' comprehension.
For example, Cruz-Lemus et al.~\cite{CruzLemus11_uml_stereotypes, Genero08_uml_stereotypes_comprehension}, Ricca et al.~\cite{Ricca10_uml_stereotypes_experiments}, and Staron et al.~\cite{Staron06_uml_stereotypes_experiments, Staron05_empirical_stereotypes} found that stereotypes (which are similar to annotations in DFDs in our experiment) increased users' efficiency and effectiveness in code comprehension.
Some publications found alternative model representations to yield better comprehension among participants in empirical experiments: Otero and Dolado~\cite{Otero05_oml_uml} reported that OPEN Modelling Language (OML) models led to faster and easier comprehension than UML diagrams, while Reinhartz-Berger and Dori~\cite{ReinhartzBerger05_opm_uml} reported Object-Process Methodology (OPM) models to be better suited than UML diagrams for modelling dynamic aspects of applications.

Bernsmed et al.~\cite{Bernsmed21} presented insights into the use of DFDs in agile teams by triangulating four studies on the adoption of DFDs.
In the studies, software engineers were confused about the structure, granularity, and what to include in the models, because no formal specification of DFDs exists.
The participants in our experiment also showed some difficulties that could be resolved by a clear definition and well-established documentation of DFDs.
Regarding DFDs' structure, Faily et al.~\cite{Faily20} argued that they should not be enriched with additional groups of model items, since their simplicity and accessibility for human users might suffer.
Instead, they proposed to use them together with other system representations.
In contrast, Sion et al. argued in a position paper~\cite{Sion20_security_threat_modeling} that using DFDs in their basic form is insufficient for threat modelling. 
Based on our findings, we argue that adding annotations to DFDs does not impede their accessibility and that security-enriched DFDs are well suited to support security analysis activities.

In conclusion, no publications were found that empirically investigate the impact of DFDs (or other model representations) on the security analysis (or related activities) of microservice applications.

\section{Conclusion}
\label{sec:conclusion}

\noindent
This paper presents the results of an empirical experiment conducted to investigate the impact of DFDs on software security analysis tasks.
DFDs are widely used for security analysis and their varied adoption indicates a high confidence in their usefulness.
To the best of our knowledge, the presented results are the first to investigate these assumptions and can confirm a positive impact of DFDs in the given context.
We found, that participants performed significantly better concerning the analysis correctness of security analysis tasks when they were provided a DFD of the analysed application.
Additionally, traceability information that links model items to artefacts in source code significantly improved their ability to provide correct evidence for their answers.
Consequently, this paper serves as a basis for future research on specific applicabilities and properties of DFDs.
Further, it can provide guidance in decisions on the adoption of model-based practices.

\section*{Acknowledgement}
\noindent This work was partly funded by the European Union's Horizon 2020 programme under grant agreement No. 952647 (AssureMOSS).

\bibliographystyle{elsarticle-num}
\bibliography{main}

\end{document}